\pdfoutput=1

\documentclass[11pt]{article}

\usepackage[]{ACL2023}

\usepackage{times}
\usepackage{latexsym}

\usepackage[T1]{fontenc}

\usepackage[utf8]{inputenc}
\usepackage{amsmath}
\usepackage{microtype}
\usepackage{multirow}
\usepackage{tabularx}
\usepackage{graphicx}
\usepackage{booktabs}
\usepackage{makecell}
\usepackage{pgfplots}
    \usepgfplotslibrary{colorbrewer}
\usepackage{adjustbox}
\usepackage{caption}
\usepackage{subcaption}
\usepackage{xcolor}
\usepackage{hyperref}
\usepackage{algorithm2e}

\SetAlCapNameFnt{\tiny}
\SetAlCapFnt{\tiny}
\usepgfplotslibrary[groupplots]
\usepackage{amssymb,graphicx,stackengine,xcolor}
\def\ucr{\scalebox{.5}{\stackinset{c}{}{c}{.2pt}{%
  \textcolor{white}{\sffamily\bfseries\tiny ?}}{%
  \rotatebox{45}{$\blacksquare$}}}}
  
\usepackage{url}
\usepackage{relsize}
\usepackage{hyperref}
\hypersetup{
  colorlinks}

\title{Discrete Prompt Optimization via Constrained Generation \\ for Zero-shot Re-ranker}

\author{Sukmin Cho
        \quad Soyeong Jeong
        \quad Jeongyeon Seo
        \quad Jong C. Park\thanks{\hspace{0.2cm}Corresponding author} \\
        School of Computing \\
        Korea Advanced Institute of Science and Technology\\ 
       \texttt{\{nelllpic,starsuzi,yena.seo,jongpark\}@kaist.ac.kr}}

\begin{document}
\maketitle
\begin{abstract}
Re-rankers, which order retrieved documents with respect to the relevance score on the given query, have gained attention for the information retrieval (IR) task. Rather than fine-tuning the pre-trained language model (PLM), the large-scale language model (LLM) is utilized as a zero-shot re-ranker with excellent results. While LLM is highly dependent on the prompts, the impact and the optimization of the prompts for the zero-shot re-ranker are not explored yet. Along with highlighting the impact of optimization on the zero-shot re-ranker, we propose a novel discrete prompt optimization method, \textbf{Co}nstrained \textbf{Prompt} generation (Co-Prompt), with the metric estimating the optimum for re-ranking. Co-Prompt guides the generated texts from PLM toward optimal prompts based on the metric without parameter update. The experimental results demonstrate that Co-Prompt leads to outstanding re-ranking performance against the baselines. Also, Co-Prompt generates more interpretable prompts for humans against other prompt optimization methods.

\end{abstract}

\setlength{\textfloatsep}{0.5em}

\section{Introduction}
Information retrieval (IR) is the task of searching for documents relevant to a given query from a large corpus. As re-ranking the fetched documents from the retriever effectively enhances the performance and the latency, recent studies have suggested several kinds of re-rankers by fine-tuning pre-trained language models (PLM)~\cite{nogueira2019passage, nogueira-etal-2020-document}. Furthermore,~\citet{sachan2022improving} show that large-scale language models (LLMs) such as GPT-3~\cite{NEURIPS2020_1457c0d6} can be exploited as a zero-shot re-ranker with the prompt describing the task. They also highlight the importance of an appropriate prompt to elicit the full performance of LLMs, rather than updating the parameters. They choose an optimal prompt among the handcrafted candidates by cross-validation. However, such a manual search for the discrete prompts is highly expensive and sub-optimal in transferability.

To resolve the issue, several methods are proposed for automatically optimizing the discrete prompt. They focus on text classification or mask-filling task while underestimating the open-ended generation~\cite{shin-etal-2020-autoprompt, gao-etal-2021-making, prasad2022grips}. Recently,~\citet{deng-etal-2022-rlprompt} address the discrete prompt optimization applicable to generation tasks with reinforcement learning by designing the reward function, which measures the generated text belonging to a discrete label. Since there are tasks that are still not aligned, requiring a continuous score of output, we aim at a prompt optimization for one of such tasks: re-ranking.

\begin{figure}
    \centering
    \includegraphics[width=0.9\columnwidth]{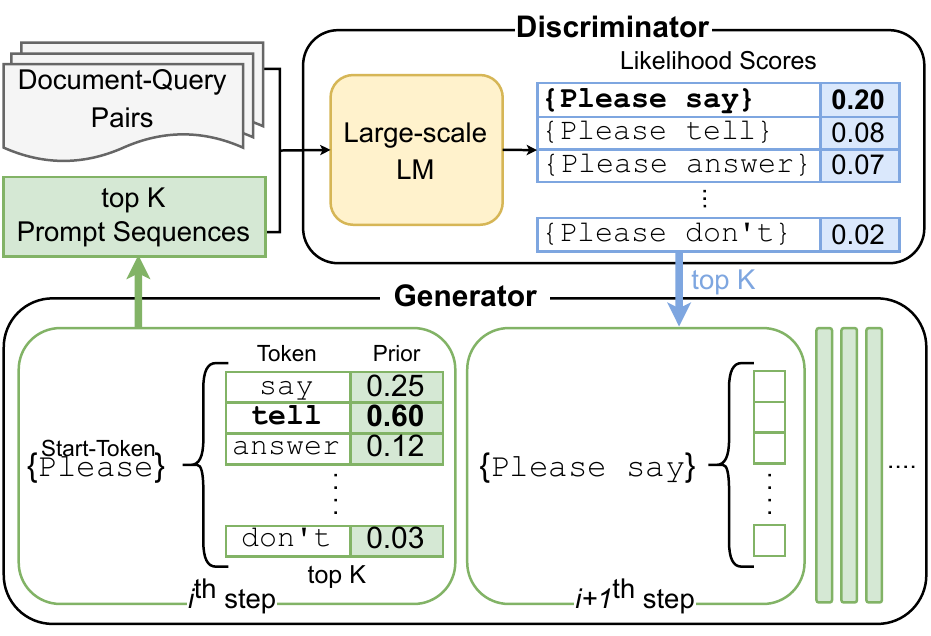}
    \caption{An overview of the constrained prompt generation process.}
    \label{fig:overview}
\end{figure}

In this paper, we propose \textbf{Co}nstrained \textbf{Prompt} generation, Co-Prompt, as left-to-right discrete prompt optimization without additional model training. By defining the metric of prompt optimum for re-ranking, we interpret the searching process of the optimal prompt as constrained generation with two modules: a zero-shot re-ranker as a discriminator and any decoder-only PLM as a generator. The discriminator calculates the likelihood (i.e., metric) that the prompt sequence is optimal for guiding an LLM to distinguish relevant documents among the large set for a given query. The generator samples the prompt tokens having a high prior from the previous prompt sequences for effectively restricting the prompt candidates for discriminator to evaluate. An overview of Co-Prompt is shown in Figure \ref{fig:overview}.

We validate our method, Co-Prompt, against other optimization baselines on two LLMs, T0~\cite{sanh2022multitask} and OPT~\cite{https://doi.org/10.48550/arxiv.2205.01068}, with two benchmark datasets, MS-MARCO~\cite{DBLP:conf/nips/NguyenRSGTMD16} and Natural Question~\cite{10.1162/tacl_a_00276}. Experimental results show that Co-Prompt consistently generates well-performing prompts regardless of LLM and dataset over the baselines. The qualitative analyses also support the interpretability of the prompts generated by Co-Prompt, similar to human language patterns.

Our contributions in this work are threefold:
\vspace{-.5em}
\begin{itemize}
    \item We highlight the impact of optimal prompt for a zero-shot re-ranker by exploiting the optimization methods.
    \vspace{-.7em}
    \item We propose Co-Prompt, a novel discrete prompt optimization via constrained generation for a zero-shot re-ranker.
    \vspace{-.7em}
    \item We experimentally show that Co-Prompt consistently guides the re-ranker well against the baselines and its output is similar to human language patterns.
\end{itemize}

\section{Related Work}
\paragraph{Document Ranking with Generative Model}
Using the generative model is one of the dominant methods for ranking the retrieved documents by defining the relevancy score as the query likelihood score~\cite{nogueira-dos-santos-etal-2020-beyond, 10.1145/3404835.3463048}. More recently,~\citet{sachan2022improving, sachan2023questions} showed that the LLM serves as either a zero-shot re-ranker or a training module of an unsupervised dense retriever. However, unlike ours, they require carefully designed manual prompts, which may have a limitation in transferability.

\paragraph{Prompt Optimization}
As prompting is considered a key variable when exploiting LLMs for various NLP tasks, finding the optimal prompt has become important to get the best performance out of the LLMs~\cite{kojima2022large, xie2022an}. Recently, the prompt optimization work has focused on discrete prompt search~\cite{shin-etal-2020-autoprompt, gao-etal-2021-making, deng-etal-2022-rlprompt} or soft prompt learning over a continuous space~\cite{liu2021gpt, qin-eisner-2021-learning,lester-etal-2021-power}. While the existing optimization methods mainly consider text classification or mask-filling task, their applicability to re-ranking is yet underexplored. In this paper, we target at optimizing discrete prompts for zero-shot re-ranker to get higher relevance scores for more relevant pairs via constrained generation.

\paragraph{Constrained Generation}
Constrained generation aims at deriving the text sequences that follow a certain constraint~\cite{keskar2019ctrl}. Utilizing a discriminator for guiding the generation toward the constraint via the Bayes' rule is one of the widely used constraint generation methods~\cite{Dathathri2020Plug, krause-etal-2021-gedi-generative, chaffin-etal-2022-ppl}. Inspired by the effectiveness of the discriminator-based method, we adopt the zero-shot re-ranker as a discriminator when generating optimal discrete prompt sequences.

\section{Method}
\subsection{Preliminaries}

An LLM re-ranks the retrieved document $d$ concerning the relevance score with a given query $q$ as the query generation score:
\begin{equation}
\small
\label{eq1}
\begin{split}
    \log P(d|q) &\propto \log P(q|d,\rho) \\
    &= \frac{1}{|q|}\sum_t \log P(q_t|q_{<t},d, \rho),
\end{split}
\end{equation} 
where $|q|$ denotes the token length of the query $q$ and $\rho$ is a natural language prompt guiding LLM to generate the query $q$. Since the prompt $\rho$ is the only controllable variable in Equation \ref{eq1}, searching for an optimal prompt is a simple yet effective way to enhance the performance of LLMs. Thus, in this work, we focus on a prompt optimization strategy.

\subsection{Constrained Prompt Generation}

We define the optimal prompt $\rho^*$ for the re-ranker which maximizes the query generation scores:
\begin{equation}
\small
\label{eq2}
\begin{split}
    \rho^* = \underset{\rho}{\arg \max}\,\, \mathbb{E}_{(d_i,q_i)\in D}[P(q_i|d_i,\rho)],
\end{split}
\end{equation}
where $D$ is the dataset for the retriever, consisting of pairs of a query and its relevant document. 

We solve the task of searching the optimal prompt $\rho^*$  for the document-query pair dataset $D$ with discriminator-based constrained generation. The generation is guided by the Bayes' rule:
\begin{equation}
\small
\begin{split}
    P(\rho_t|D, \rho_{1:t-1}) &\propto P_{M_D}(D_s|\rho_{1:t})P_{M_G}(\rho_t| \rho_{1:t-1}),
\end{split}
\label{eq3}
\end{equation}
where $M_D$ is a zero-shot re-ranker serving as a discriminator, $M_G$ is a decoder-only PLM as a generator, and $D_s$ is a dataset sampled from $D$. 

\paragraph{Discriminator} The discriminator $M_D$ measures how effectively the prompt sequence $\rho_{1:t}$ guides the zero-shot re-ranker to generate the query from the given document by computing the likelihood $P_{M_D}(D_s|\rho)$, defined as the expectation of relevance score between document-query pairs $(q_i,d_i)$ of the sampled dataset $D_s$ with the prompt $\rho$:
\begin{equation}
\small
\begin{split}
    P_{M_D}(D_s|\rho) = \mathbb{E}_{(d_i,q_i)\in D_s}[P_{M_D}(q_i|d_i,\rho)].
\end{split}
\label{eq4}
\end{equation}
We use this likelihood as the metric for prompt optimum. The other option of $P_{M_D}$ is shown in Appendix \ref{Likelihood}.

\RestyleAlgo{ruled}
\SetKwInput{kwInit}{Require}
\begin{algorithm}[t]
    \caption{Co-Prompt: a beam search-based prompt generation algorithm with a discriminator and a generator. $D_s$: document-query pairs, $B$: beam width, $L$: maximum prompt length, $N$: the number of final prompts, $\mathcal{V}$: vocabulary set}\label{alg:co-prompt}
    \tiny
    \kwInit{$D_s, B, L, \mathcal{V}$}
    \Begin{
        $P_1 \gets \{\textnormal{Start-Token}\}$ \\
        \For{$t=1,\dots,L$}{
            $P_{t+1} \gets \emptyset $ \\
            
            \ForEach{$\rho_{1:t} \in P_{t}$}{
                $S_{t+1} \gets \underset{K=B,\rho_{t+1}\in\mathcal{V}}{topK}\,P_{M_G}(\rho_{t+1}|\rho_{1:t})$ \\
                $P_{t+1} \gets P_{t+1} \cup \{\rho_{1:t+1}|\rho_{1:t} \oplus \rho_{t+1} \in S_{t+1}\}$ 
            }
            $P_{t+1} \gets \underset{K=B,\rho_{1:t+1}\in P_{t+1}}{topK}\,P_{M_D}(D_s|\rho_{1:t+1})$
        }
        $P \gets \cup_{t\in [1,L]} P_t$ \\ 
        $R \gets \underset{K=N,\rho\in P}{topK}\, P_{M_D}(D_s|\rho)$ \\
        \Return $R$
    }
\end{algorithm}

\paragraph{Generator} The generator $M_G$ samples the pool of prompts to be evaluated by a discriminator since computing Equation \ref{eq3} of all possible tokens in the vocabulary requires a prohibitively high computational cost. The decoder-only PLM is exploited to sample prompt tokens $\rho_{t}$ having a high prior $P_{M_G}(\rho_t|\rho_{1:t-1})$ in a zero-shot manner.

We combine these modules to optimize the prompt by iteratively performing two steps: candidate generation and evaluation. We choose to use a beam search as a decoding strategy for left-to-right prompt generation. The detailed steps of the decoding strategy are shown in Algorithm \ref{alg:co-prompt}.

\begin{table}[t]
    \centering
    \tiny
    \begin{tabularx}{\columnwidth}{ >{\raggedright\arraybackslash\hsize=3.8\hsize}X | >{\centering\arraybackslash\hsize=0.65\hsize} X >{\centering\arraybackslash\hsize=0.65\hsize} X | >
    {\centering\arraybackslash\hsize=0.65\hsize} X >{\centering\arraybackslash\hsize=0.65\hsize} X | >{\centering\arraybackslash\hsize=0.65\hsize} X >{\centering\arraybackslash\hsize=0.65\hsize} X | >{\centering\arraybackslash\hsize=0.65\hsize} X >{\centering\arraybackslash\hsize=0.65\hsize} X}
        \toprule 
         & \multicolumn{4}{c|}{\textbf{NQ}} & \multicolumn{4}{c}{\textbf{MS-MARCO}} \\
         & \multicolumn{2}{c|}{\textbf{BM25}} & \multicolumn{2}{c|}{\textbf{DPR}} & \multicolumn{2}{c|}{\textbf{BM25}} & \multicolumn{2}{c}{\textbf{DPR}} \\ \cmidrule{2-9}
        \textbf{ACC@$k$($\rightarrow$)} & 20 & 100 & 20 & 100 & 20 & 100 & 20 & 100 \\ \midrule
        Only Retriever & 62.9 & 78.3 & 79.2 & 85.7 & 48.0 & 66.7 & 37.5 & 55.5 \\ \midrule 
        \multicolumn{9}{c}{\textit{T0-3B Re-ranker}} \\ \midrule
        Null Prompt & 73.1 & 82.8 & 78.5 & 86.6 & 53.2 & 72.7 & 51.5 & 68.0  \\ 
        P-tuning & 72.8 & 82.7 & 79.1 & 87.0 & 54.1 & 72.5 & 52.5 & 68.2 \\
        RL Prompt & 74.7 & \underline{83.4} & 79.9  & 87.4  & \underline{60.9} & 77.4 & 57.1 & 71.2  \\ 
        Manual Prompt & \textbf{75.7} & \textbf{83.8} & \textbf{81.3} & \textbf{87.8} & 60.6 & \underline{77.9} & \underline{57.7} & \textbf{72.0} \\ 
        Co-Prompt (Ours) & \underline{75.0} & \textbf{83.8} & \underline{80.4} & \underline{87.7} & \textbf{61.9} & \textbf{78.0} & \textbf{58.0} & \underline{71.7} \\ \midrule
        \multicolumn{9}{c}{\textit{OPT-2.7B Re-ranker}} \\ \midrule
        Null Prompt & 70.5 & 81.9 & 76.3  & 86.1  & 50.4 & 71.7 & 50.1 & 68.1  \\ 
        P-tuning & 71.2 & 82.8 & 78.3 & \underline{87.5} & 56.5 & 75.5 & 54.6 & 69.9 \\
        RL Prompt & 72.5 & 82.9 & \underline{79.1} & 87.4 & \underline{59.2} & \underline{76.7} & \underline{56.3} & \underline{71.1}  \\
        Manual Prompt & \underline{73.1} & \underline{83.3} & 78.9 & 87.2 & 55.3 & 74.6 & 54.3 & 70.1 \\ 
        Co-Prompt (Ours) & \textbf{75.2} & \textbf{84.1} & \textbf{80.2} & \textbf{88.1} & \textbf{59.3} & \textbf{77.2} & \textbf{56.4} & \textbf{71.3} \\
        \bottomrule
    \end{tabularx}
    \vspace{-0.5em}
    \caption[Main table 1]
    {ACC@$k$ of the re-ranked result with the prompts when $k$ is 20 and 100. The best scores are marked in \textbf{bold}, and the next ones are \underline{underlined}.}
    \label{tab:overall_result}
\end{table}

\begin{table*}[t]
    \centering
    \tiny
    \begin{tabularx}{\textwidth}{
    >{\raggedright\arraybackslash\hsize=0.45\hsize}X
    >{\raggedright\arraybackslash\hsize=0.6\hsize}X  |
    >{\raggedright\arraybackslash\hsize=2.2\hsize} X | 
    >{\centering\arraybackslash\hsize=.25\hsize} X |
    >{\raggedright\arraybackslash\hsize=2.2\hsize} X | 
    >{\centering\arraybackslash\hsize=.25\hsize} X
    }
    \toprule
    \textbf{Retriever} & \multirow{2}{*}{\textbf{Prompt}} &  \multicolumn{2}{c|}{\textbf{MS-MARCO}} &  \multicolumn{2}{c}{\textbf{NQ}}   \\ 
    \textbf{\textbackslash Re-ranker} &  & \textbf{Instruction Prompt} & \textbf{nDCG} & \textbf{Instruction Prompt} & \textbf{nDCG} \\ \midrule
    BM25 & - & - & 25.2 & - & 20.2 \\ \midrule
    \multirow{7}{*}{OPT} & Manual Prompt & "\textit{Please write a question based on this passage}" & 28.7 & "\textit{Please write a question based on this passage}" & 27.9 \\ 
     & RL-Prompt &  "\textit{questions answers key question defining}" & 31.5 & "\textit{ poll trivia trivia wondered asking}" & 27.2 \\ \cmidrule{2-6}
     & \multirow{5}{*}{Co-Prompt} & "\textit{Please tell that\ucr \space is the first question asked on Google for}" & \textbf{31.9} & "\textit{Please post your question again when its not just about}" & 30.6 \\
      & & "\textit{Score! What are all 3 things, the first is}" & 30.2 & "\textit{Score the top 5 things on this sub reddit for}" & 29.3 \\
       &  &"\textit{This looks like the same as every "what are the}" & 30.5 & "\textit{This post should be titled as}" & \textbf{31.2} \\
        & & "\textit{What are some common questions asked on the internet about}" & 30.3 & "\textit{How do i find the name on google, and}" & 29.1 \\
    \bottomrule
    \end{tabularx} 
    \vspace{-0.5em}
    \caption{Comparison of different discrete prompts and evaluation on the top-20 documents retrieved by BM25. The best results of each re-ranker are marked in \textbf{bold}.}
    \label{tab:generator}
    \vspace{-2em}
\end{table*}

\begin{table}[ht]
    \centering
    \tiny
    \begin{tabularx}{\columnwidth}{
    >{\raggedright\arraybackslash\hsize=0.6\hsize}X
    >{\raggedright\arraybackslash\hsize=0.6\hsize}X  |
    >{\centering\arraybackslash\hsize=.7\hsize} X
    >{\centering\arraybackslash\hsize=.7\hsize} X 
    }
    \toprule
      \textbf{Retriever} & \textbf{Prompt} & \multicolumn{2}{c}{\textbf{MSMARCO}}   \\
      \textbf{\textbackslash Re-ranker} & \textbf{\textbackslash Generator} & \textbf{nDCG@20} & \textbf{nDCG@100} \\ \midrule
      BM25 & - &  22.84 & 28.70  \\ \midrule
      \multirow{3}*{T0} & GPT2-Base &  30.76 & 36.44  \\  
     & GPT2-Large & \textbf{31.11} & \textbf{36.79}  \\ 
     & GPT2-XL & 29.86 & 35.71 \\ 
     \bottomrule
    \end{tabularx}
    \vspace{-0.5em}
    \caption{Comparison between the prompts from the different generators. The best results are marked in \textbf{bold}.}
    \label{tab:generator_3}
\end{table}
\section{Experimental Setups}
We describe the experimental setups for validating the performance of the prompts. Our code is publicly available at \href{https://github.com/zomss/Co-Prompt}{github.com/zomss/Co-Prompt}.

\paragraph{Datasets} We employ two information retrieval datasets: \textbf{1) MS-MARCO}~\cite{DBLP:conf/nips/NguyenRSGTMD16}, collected from the Bing search logs, and \textbf{2) Natural Question (NQ,}~\citet{10.1162/tacl_a_00276}\textbf{)}, fetched from Google search engines. We only use the document data of the dataset for evaluation. More information is shown in Appendix \ref{Datasets}.

\paragraph{Evaluation Metrics}\label{sec:metric} We evaluate the results by two metrics, ACC and nDCG. \textbf{1) ACC} is the percentage of the relevant documents in the total retrieved ones. \textbf{2) nDCG}, normalized discounted cumulative gain, reflects that the more relevant documents should record higher ranks.

\paragraph{Retriever \& Re-ranker} We select two widely used sparse and dense retrievers as our retrievers, which are \textbf{1) BM25}~\cite{INR-019} and \textbf{2) DPR}~\cite{karpukhin-etal-2020-dense}, respectively. For the zero-shot re-ranker, we use \textbf{1) T0}~\cite{sanh2022multitask} and \textbf{2) OPT}~\cite{https://doi.org/10.48550/arxiv.2205.01068}. We describe more detailed information in Appendix \ref{Retrievers} and \ref{Re-rankers}.

\paragraph{Prompt Baselines} We compare Co-Prompt against four baselines: \textbf{1) Null Prompt} is an empty prompt without any token.~\textbf{2) P-Tuning} is a soft prompt optimization method that yields prompt embeddings from the prompt encoder~\cite{liu2021gpt}.~\textbf{3) RL-Prompt} is a discrete prompt optimization method by training policy network~\cite{deng-etal-2022-rlprompt}. Note that we modify RL-Prompt and P-Tuning applicable for the re-ranking task.~\textbf{4) Manual Prompt}, suggested at~\citet{sachan2022improving}, is given as "\textit{Please write a question based on this passage}", following the assumption that it is one of the best prompts that humans can find. Last,~\textbf{5) Co-Prompt}, our proposed method, is a discrete prompt optimization method in a left-to-right zero-shot generation. Note that the implementation detail of baselines is shown in Appendix \ref{Baselines}.

\paragraph{Implementation Details} The discriminator $M_D$ is the same model as the zero-shot re-ranker. Since the generator $M_G$ should be a decoder-only model, in the case of T0, GPT2-Large~\cite{radford2019language} is utilized as the generator. OPT, a decoder-only model, is used as both the discriminator and generator. We use the start token as "Please" for direct comparison with the manual prompt and fix the beam width $B$ as 10 and the maximum prompt length $L$ as 10 in our experiment. 

\paragraph{Environment} We conduct all experiments including prompt searching and document re-ranking on V100 32GB GPUs. We use BEIR~\cite{thakur2021beir} framework\footnote{\url{http://beir.ai/}} for re-ranked result evaluation and passage retrieval datasets. Also, the retrievers, BM25 and DPR, are from the same framework. We employ T0 and OPT with 3B and 2.7B parameters each for the discriminator and the re-ranker publicly open on the Huggingface model hub\footnote{\url{https://huggingface.co/models}}~\cite{wolf-etal-2020-transformers}. 

\section{Result}
In this section, we show the overall results of our method, Co-Prompt, with a detailed analysis. 

\paragraph{Overall Results} As shown in Table \ref{tab:overall_result}, Co-prompt consistently shows a robust performance gain in all scenarios, regardless of LLM, the dataset, and the retriever. Specifically, Co-Prompt, applied to OPT, achieves better results than the other methods. This indicates that the prompts generated by our method are more appropriate to play the role of an instruction to guide LLMs against other prompt optimization methods. More detailed results of re-ranked performance with various metrics are shown in Appendix \ref{sec:gen_prompts}. 

\paragraph{Impact of Start Tokens}
 We exploit other options of start token such as "Score" and "This" as shown in Table \ref{tab:generator}. Regardless of the start tokens, Co-Prompt consistently generates prompts eliciting the performance of LLM efficiently. However, we observe that finding the optimal start token for the dataset is important to achieve better results. 

\paragraph{Impact of Generator} 
As shown in Table \ref{tab:generator_3}, even if different generators are used, the generated prompts by different generators guide the zero-shot re-ranker efficiently. Still, the differences in performance are caused by a vocabulary mismatch between the two modules. We see that, although our method does not vary significantly in performance to the generator, a more suitable generator may be necessary for better results.

\begin{figure}
    \vspace{-1em}
    \centering
    \includegraphics[width=\columnwidth]{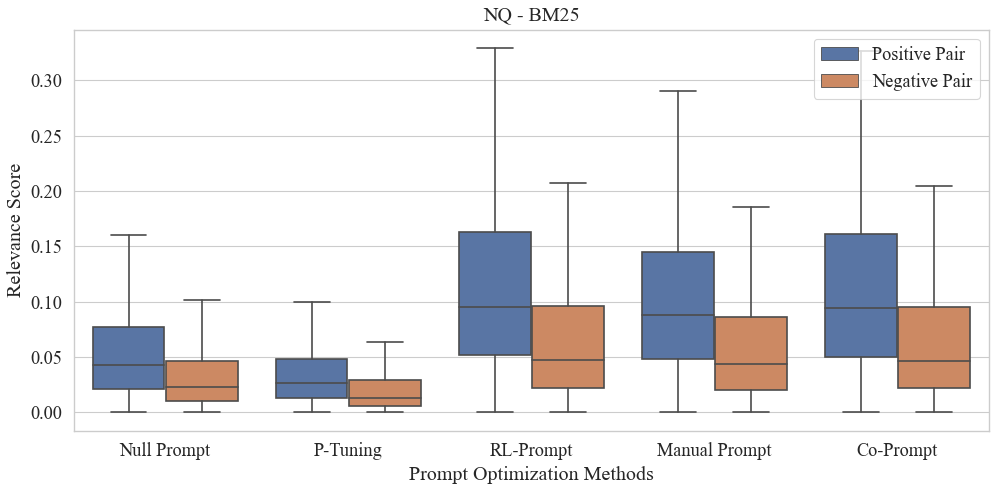}
    \vspace{-1em}
    \caption{Distributions of relevance scores between document-query pairs. The positive pairs mean relevant ones and negative pairs are irrelevant.}
    \label{fig:distribution}
\end{figure}

\paragraph{Relevance Score} We analyze the distributions of relevance scores between positive or negative document-query pairs. As the negative documents for a given query are retrieved from BM25, the negative ones are related to the query but unable to directly find the answer. As shown in Figure \ref{fig:distribution}, we point out that the distribution difference exists between pairs despite some overlap. Also, LLM can distinguish which pair is positive, even without a prompt. However, we observe that the effect of discrete prompt optimization on the zero-shot re-ranker is in the direction of increasing the mean and variance of the relevance score.

\paragraph{Case Study of Prompts} Table \ref{tab:generator} shows the discrete prompts generated by our method and discrete prompt baselines when exploiting OPT as a re-ranker. While the prompts from the RL-prompt are ungrammatical gibberish close to a random word sequence, our method, Co-Prompt, generates interpretable prompts for humans, following human language patterns, and surpasses the performance of the other discrete prompts. Also, the word `\textit{question}', one of the keywords describing the task, is included in the prompts from Co-Prompt regardless of the datasets. This implies that the prompts from our method can provide a natural user interface to improve human understanding of how LLMs work. See Appendix \ref{sec:gen_prompts} for more examples of Co-Prompt.

\section{Conclusion}
In this paper, we propose Co-Prompt, left-to-right prompt optimization for zero-shot re-ranker via constrained generation. Co-Prompt effectively restricts prompt candidates and evaluates the optimum of these prompts without any parameter updates. We experimentally show that our method achieves consistently outperforming performance across all experiments. Also, the impact of prompt optimization including baselines on the zero-shot re-ranker highlights its importance. We also present an interesting outcome in that the optimal prompt is interpretable for human. For future work, we plan to expand our method to other open-ended generation tasks using LLMs.

\section*{Limitations}
As shown in Table \ref{tab:overall_result}, our method is experimentally demonstrated to be effective in two LLMs. However, OPT, a decoder-only model, is more suitable for the prompts generated by Co-Prompt. This seems to be because T0, the encoder-decoder model, requires a separate generator such as GPT-2. The performance of prompts may vary depending on the generator involved in the vocabulary and training process. Also, there is a trade-off between search time and performance. While increasing the beam size and the number of document-query pairs enhances the probability of finding a more optimal prompt, it makes the search time proportionally longer. 

\section*{Ethics Statement}
Our work contributes to enhancing the retrieval performance of a zero-shot re-ranker by optimizing the discrete prompt via constrained generation. We are keenly aware of the possibility of offensive or upsetting prompts caused by bias of the generator itself even though there were no such prompts in our experiments. Because there is no additional training for prompt optimization, our method has difficulty removing the bias of the language model itself. As studies on reducing the bias of language models or filtering out inappropriate expressions in texts are being actively conducted, these problems are expected to be sufficiently resolved in the future.
\section*{Acknoledgements}
This work was supported by Institute for Information and communications Technology Promotion (IITP) grant funded by the Korea government (No. 2018-0-00582, Prediction and augmentation of the credibility distribution via linguistic analysis and automated evidence document collection).
\bibliography{custom}
\bibliographystyle{acl_natbib}

\clearpage
\appendix

\section{Implementation Details}\label{Impl}
\begin{table*}[h]
    \centering
    \tiny
    \begin{tabularx}{\textwidth}{
    >{\raggedright\arraybackslash\hsize=0.5\hsize}X
    >{\raggedright\arraybackslash\hsize=0.6\hsize}X  |
    >{\raggedright\arraybackslash\hsize=2.8\hsize} X | 
    >{\centering\arraybackslash\hsize=.4\hsize} X
    >{\centering\arraybackslash\hsize=.4\hsize} X
    >{\centering\arraybackslash\hsize=.4\hsize} X
    >{\centering\arraybackslash\hsize=.4\hsize} X
    }
    \toprule
      \textbf{Retriever} & \textbf{Prompt} & \multirow{2}{*}{\textbf{Instruction Prompt}} & \multicolumn{4}{c}{\textbf{MS-MARCO}}   \\
      \textbf{\textbackslash Re-ranker} & \textbf{\textbackslash Generator} & & nDCG@20 & nDCG@100 & MAP@20 & MAP@100  \\ \midrule
      BM25 & - & - & 22.84 & 28.70 & 18.69 & 65.78 \\ \midrule
      \multirow{5}*{T0} & Manual Prompt & "\textit{Please write a question based on this passage.}" & 30.31 & 36.13 & 24.03 & 25.22 \\  \cmidrule{2-7}
      & GPT2-Base & "\textit{Please and tell me why, what, how,}" & 30.76 & 36.44 & 24.54 & 25.70  \\  
     & GPT2-Large & "\textit{Please send me some info on why or in detail}" & \textbf{31.11} & \textbf{36.79} & \textbf{24.82} & \textbf{25.99} \\ 
     & GPT2-XL & "\textit{Please enter the message content, such\textbackslash n and\textbackslash n}" & 29.86 & 35.71 & 23.99 & 25.17  \\ 
     \bottomrule
    \end{tabularx}
    \caption{Comparison of the prompts from the different generators and evaluation on the document set retrieved from MS-MARCO by BM25. The best results of each metric are marked in \textbf{bold}.}
    \label{tab:generator_2}
\end{table*}
\begin{table}[t]
    \centering
    \tiny
    \begin{tabularx}{\columnwidth}{
    >{\raggedright\arraybackslash\hsize=0.1\hsize}X
    >{\raggedright\arraybackslash\hsize=1.5\hsize}X |
    >{\centering\arraybackslash\hsize=.7\hsize} X
    >{\centering\arraybackslash\hsize=.7\hsize} X |
    >{\centering\arraybackslash\hsize=.7\hsize} X
    >{\centering\arraybackslash\hsize=.7\hsize} X
    }
    \toprule
    \multicolumn{2}{l|}{\textbf{Retriever}} &  \multicolumn{2}{c|}{\textbf{NQ}} &  \multicolumn{2}{c}{\textbf{MS-MARCO}}   \\ 
    \multicolumn{2}{l|}{\textbf{\textbackslash Re-ranker}} & \textbf{ACC@20} & \textbf{ACC@100} & \textbf{ACC@20} & \textbf{ACC@100} \\ \midrule
    \multicolumn{2}{l|}{BM25}  & 62.9 & 78.3 & 48.0 & 66.7 \\ 
    \multirow{2}{*}{T0} & + Base Metric & 75.0 & 83.8 & 61.9 & 78.0 \\
    & + Contrastive Metric & 76.2 & 83.8 & 59.6 & 76.2 \\ 
    \multirow{2}{*}{OPT} & + Base Metric & 75.2 & 84.1 & 59.3 & 77.2 \\
    & + Contrastive Metric & 74.4 & 84.0 & 57.7 & 75.7 \\ \midrule
    \multicolumn{2}{l|}{DPR} & 79.2 & 85.7 & 37.5 & 55.5 \\ 
    \multirow{2}{*}{T0} & + Base Metric & 80.4 & 87.7 & 58.0 & 71.7 \\
    & + Contrastive Metric & 80.6 & 87.9 & 56.4 & 70.8 \\
    \multirow{2}{*}{OPT} & + Base Metric & 80.2 & 88.1 & 56.4 & 71.3 \\
    & + Contrastive Metric & 80.2 & 87.9 & 53.3 & 68.9 \\
    \bottomrule
    \end{tabularx} 
    \caption{Comparison between two options of likelihood at the ACC-$k$ accuracy.}
    \label{tab:likelihood}
\end{table}

\subsection{Datasets}\label{Datasets}

We employ two information retrieval datasets for evaluating the performance of the zero-shot re-ranker with the prompts. \textbf{1) MS-MARCO}~\cite{DBLP:conf/nips/NguyenRSGTMD16} contains about 8M passages and 6,980 queries in development split collected from the Bing search logs. Because of the diversity of topics and contents with the large training set, recent work exploits MS-MARCO for retriever training~\cite{nogueira2019passage, qu-etal-2021-rocketqa}. \textbf{2) Natural Question (NQ,}~\citet{10.1162/tacl_a_00276}\textbf{)} contains about 2M passages of Wikipedia articles and 3,452 queries in test split collected from Google search engines. Also, NQ, one of the popular open-domain question datasets, is exploited as training data of dense retrievers~\cite{karpukhin-etal-2020-dense}. Both datasets are the benchmarks for evaluating information retriever systems~\cite{thakur2021beir}. Only 1,500 document-query pairs from MS-MARCO test split and NQ development split each are utilized for the prompt optimization. 

\subsection{Metrics}
As mentioned in Section \ref{sec:metric}, we employ two metrics, \textbf{1) ACC} and \textbf{2) nDCG}. In addition, we use one more metric. \textbf{3) MAP} is the mean average precision of the relevant documents' ranks for a given query.

\subsection{Retrievers}\label{Retrievers}

We use two types of retrievers, sparse and dense retrievers, for retrieving documents re-ranked by LLMs. \textbf{1) BM25}~\cite{INR-019} is a representative sparse retriever computing the relevancy score between a document and a query based on term frequency and inverse document frequency. BM25 has been widely employed because of its fast speed and effective performance. \textbf{2) DPR}~\cite{karpukhin-etal-2020-dense} interprets training dense retrieval as metric learning problems. The bi-encoder initialized with BERT~\cite{devlin-etal-2019-bert} is trained with contrastive learning exploiting positive and negative passages for a given query. It shows outperforming results over traditional sparse retrievers.

\subsection{Zero-shot Re-rankers}\label{Re-rankers}

We employ two LLMs, T0 and OPT, as re-rankers with the prompt. 
\textbf{1) T0}, one of the T5 series ~\cite{JMLR:v21:20-074}, consists of transformer encoder-decoder layers. The models are fine-tuned versions of T5 for multi-task learning with prompted datasets. \textbf{2) OPT}, a publicly open model, consists of decoder-only transformer layers. Its performance is comparable to those of GPT-3 models. We exploit OPT instead of GPT-3 due to academic budget. 

The template is needed when trasmitting a document, a prompt and a query to zero-shot re-ranker together. Following the template setting of UPR, the template used in the experiments is "Passage: \{document\} \{delimiter\} \{prompt\} \{delimiter\} \{query\}". The delimiters used in the experiments are " " for T0 and "\textbackslash n" for OPT.

\subsection{Baselines}\label{Baselines}

\paragraph{Manual Prompt}~\citet{sachan2022improving} not only proposed unsupervised passage re-ranker exploiting LLMs but also carefully selected the optimal prompt among handcrafted candidates validated by the re-ranked result at BM25 passages of NQ development set. The manually optimized prompt "\textit{Please write a question based on this passage}" effectively guides zero-shot re-rankers to generate the query corresponding to the document.


\paragraph{P-tuning}~\citet{liu2021gpt} proposed P-tuning\footnote{\url{https://github.com/THUDM/P-tuning}}, generating soft prompts (i.e., continuous prompt embeddings), not discrete ones. They employed the prompt encoder consisting of long-short term memory layers trained to return the optimal soft prompts for the task. While the method mainly focuses on the text classification task, we define the loss objective as query generation log-likelihood for application to re-ranking. The prompt encoder is trained with document-query pairs for 10 epochs to generate 10-length soft prompts.

\paragraph{RL-Prompt}~\citet{deng-etal-2022-rlprompt} proposed discrete prompt generation, applicable to open-ended generation tasks, with reinforcement learning. They validated the method applicable to text style transfer, one of open-ended text generation techniques. In order to align to the re-ranking task, we define the reward for the policy network as query generation log-likelihood from the document and the prompt. Following the setting mentioned in RL-Prompt\footnote{\url{https://github.com/mingkaid/rl-prompt}}, the 5-token length prompt is created through 12,000 training steps with a policy network model.

\section{Analysis}

\subsection{Likelihood $P_{M_D}(D_s|\rho_{1:t})$}\label{Likelihood}

In this section, we call the likelihood proposed in Equation \ref{eq4} as the base metric. We consider the other option of likelihood $P_{M_D}(D_s|\rho_{1:t})$ in contrastive manner and also show the compared result with base metric in Table \ref{tab:likelihood}.

\paragraph{Contrastive Measurement} The query generation score should be high for positive document-query pairs $D_s^+$ and low for negative pairs $D_s^-$. In a contrastive manner, the likelihood exploits the contrast between $P_{base}(D_s^+|\rho)$ and $P_{base}(D_s^-|\rho)$ as follows:

\begin{equation}
\small
\begin{split}
    P_{cont}(D_s|\rho) = \frac{P_{base}(D_s^+|\rho)}{P_{base}(D_s^+|\rho) + P_{base}(D_s^-|\rho)}
\end{split}
\end{equation}

As shown in Table \ref{tab:likelihood}, base metric gains a certain level of performance regardless of the dataset and LLM, whereas contrastive metric shows inferior performance over MS-MARCO.

\subsection{Impact of Generator}

We show more detailed results of the prompts from the different generators in table \ref{tab:generator_2}. While the generated prompts follow human language patterns, there are some differences in used words. 

\subsection{Detailed Results}\label{sec:gen_prompts}

\begin{table*}[h]
    \centering
    \tiny
    \begin{tabularx}{\textwidth}{
    >{\raggedright\arraybackslash\hsize=0.3\hsize} X  
    >{\raggedright\arraybackslash\hsize=0.6\hsize} X |  
    >{\raggedright\arraybackslash\hsize=3.0\hsize} X | 
    >{\centering\arraybackslash\hsize=.4\hsize} X
    >{\centering\arraybackslash\hsize=.4\hsize} X
    >{\centering\arraybackslash\hsize=.4\hsize} X
    >{\centering\arraybackslash\hsize=.4\hsize} X
    >{\centering\arraybackslash\hsize=.4\hsize} X
    >{\centering\arraybackslash\hsize=.4\hsize} X 
    }
    \toprule
      \multicolumn{2}{l|}{\textbf{Retriever}} & \multirow{2}{*}{\textbf{Instruction Prompt}} & \multicolumn{6}{c}{\textbf{NQ}}  \\
      \multicolumn{2}{l|}{\textbf{\textbackslash Re-ranker}} & & ACC@20 & ACC@100 & nDCG@20 & nDCG@100 & MAP@20 & MAP@100  \\ \midrule
      \multicolumn{2}{l|}{BM25} & - & 62.9 & 78.3 & 20.2 & 23.9 & 7.8 & 9.9  \\ \midrule
      \multirow{8}{*}{T0} & Null & "" & 73.1 & 82.8 & 27.8 & 32.1 & 12.9 & 16.0 \\ 
        & P-Tuning & - & 72.9  & 82.8 & 27.9 & 32.2 & 12.8 & 16.0  \\
      & RL-Prompt & "\textit{ poll question question question knows}" & 74.7 & 83.4 & 30.4 & 34.6 & 14.4 & 17.9  \\
       & Manual & "\textit{Please write a question based on this passage}" & \textbf{75.7} & \textbf{83.8} & \textbf{32.5} & \textbf{36.6} & \textbf{15.9} & \textbf{19.7}  \\ 
       & \multirow{3}{*}{Co-Prompt} & "\textit{Please try and find out the answer by asking questions like}" & 75.0 & \textbf{83.8} & 30.9 & 35.1 & 14.8 & 18.4  \\
       & & "\textit{Please try and find out the answer by asking questions below}" & 75.1 & 83.5 & \underline{31.0} & \underline{35.2} & \underline{15.0} & \underline{18.5} \\
       & & "\textit{Please try and find out the answer by asking questions} & \underline{75.3} & \underline{83.7} & \underline{31.0} & 35.1 & 14.9 & 18.4  \\  \midrule
      \multirow{8}{*}{OPT} & Null & "" & 70.5 & 81.9 & 25.1 & 29.8 & 11.1 & 14.0  \\ 
       & P-Tuning & - & 71.2 & 82.9 & 27.2 & 32.1 & 12.5 & 15.9  \\
       & RL-Prompt & "\textit{ poll trivia trivia wondered asking}" & 72.5 & 82.9 & 27.2 & 31.7 & 12.3 & 15.5  \\
       & Manual & "\textit{Please write a question based on this passage}" & 73.2 & 83.3 & 27.9 & 32.5 & 12.9 & 16.2  \\ 
       & \multirow{3}{*}{Co-Prompt} & "\textit{Please post your question again when its not about the}" & \underline{75.2} & \textbf{84.1} & \underline{30.4} & \underline{34.9} & \underline{14.4} & \underline{17.9}  \\
       & & "\textit{Please post your question again when its not just about}" & \textbf{75.5} & \textbf{84.1} & \textbf{30.6} & \textbf{35.1} & \textbf{14.7} & \textbf{18.2}   \\
       & & "\textit{Please post your question again after doing research about} & 74.5 & \underline{83.9} & 29.6 & 34.2 & 14.0 & 17.4  \\  \midrule \midrule
      \multicolumn{2}{l|}{DPR} & - & 79.2 & 85.7 & 34.0 & 35.2 & 17.9 & 19.8  \\ \midrule
      \multirow{8}{*}{T0} & Null & "" & 78.5 & 86.6 & 31.4 & 34.9 & 15.9 & 18.6  \\ 
        & P-Tuning & - & 79.1 & 87.0 & 32.1 & 35.4 & 16.1 & 19.0  \\
      & RL-Prompt & "\textit{ poll question question question knows}" & 79.9 & 87.4 & 34.1 & 37.5 & 17.4 & 20.5 \\
       & Manual & "\textit{Please write a question based on this passage}" & \textbf{81.4} & \textbf{87.8} & \textbf{36.6} & \textbf{39.7} & \textbf{19.1} & \textbf{22.5}  \\ 
       & \multirow{3}{*}{Co-Prompt} & "\textit{Please try and find out the answer by asking questions like}" & \underline{80.4} & \underline{87.7} & 34.5 & 38.0 & 17.8 & 21.0   \\
       & & "\textit{Please try and find out the answer by asking questions below}" & 80.2 & 87.6 & \underline{34.8} & \underline{38.2} & \underline{17.9} & \underline{21.2} \\
       & & "\textit{Please try and find out the answer by asking questions} & 80.2 & 87.6 & \underline{34.8} & 38.1 & \underline{17.9} & 21.1   \\  \midrule
      \multirow{8}{*}{OPT} & Null & "" & 76.3 & 86.1 & 28.8 & 32.8 & 13.8 & 16.5  \\ 
        & P-Tuning & - & 78.2 & 87.5 & 31.8 & 36.1 & 15.9 & 19.2  \\
       & Manual & "\textit{Please write a question based on this passage}" & 78.9 & 87.5 & 32.0 & 35.8 & 16.0 & 19.0 \\
       & RL-Prompt & "\textit{ poll trivia trivia wondered asking}" & 79.1 & 87.0 & 31.6 & 35.2 & 15.7 & 18.6 \\
       & Manual & "\textit{Please write a question based on this passage}" & 78.9 & 87.5 & 32.0 & 35.8 & 16.0 & 19.0 \\ 
       & \multirow{3}{*}{Co-Prompt} & "\textit{Please post your question again when its not about the}" & \textbf{80.2} & \underline{88.1} & \underline{34.1} & \underline{37.8} & \underline{17.3} & \underline{20.5}   \\
       & & "\textit{Please post your question again when its not just about}"& \textbf{80.2} & 88.0 & \textbf{34.6} & \textbf{38.2} & \textbf{17.8} & \textbf{21.0}     \\
       & & "\textit{Please post your question again after doing research about} & \underline{80.1} & \textbf{88.3} & 33.6 & 37.6 & 17.1 & 20.3   \\ 
     \bottomrule \\
    \toprule
      \multicolumn{2}{l|}{\textbf{Retriever}} & \multirow{2}{*}{\textbf{Instruction Prompt}} & \multicolumn{6}{c}{\textbf{MS-MARCO}}  \\
      \multicolumn{2}{l|}{\textbf{\textbackslash Re-ranker}} & & ACC@20 & ACC@100 & nDCG@20 & nDCG@100 & MAP@20 & MAP@100  \\ \midrule
      \multicolumn{2}{l|}{BM25} & - & 48.0 & 66.7 & 25.2 & 28.7 & 18.7 & 19.2  \\ \midrule
      \multirow{8}{*}{T0} & Null & "" & 53.2 & 72.7 & 27.5 & 31.2 & 20.2 & 20.7  \\ 
             & P-Tuning & - & 54.1 & 72.5 & 28.5 & 31.9 & 21.1 & 21.6  \\
      & RL-Prompt & "\textit{ question meaning difference meaning reality}" & 60.9 & 77.4 & 33.1  & 35.2 & 25.1 & 25.4  \\
       & Manual & "\textit{Please write a question based on this passage}" & 60.6 & \underline{77.9} & 32.8 & 36.1 & 24.8 & 25.2  \\ 
       & \multirow{3}{*}{Co-Prompt} & "\textit{Please send me some info on why or in detail,}" & \textbf{61.9} & \textbf{78.0} & \textbf{33.7} & \textbf{36.8} & \textbf{25.5} & \textbf{26.0}  \\
       & & "\textit{Please send me some info on why or in detail about}" & \underline{61.2} & 77.8 & \underline{33.4} & \underline{36.6} & \underline{25.4} & \underline{25.9}  \\
       &  & "\textit{Please send me some info on why or in detail on}" & 61.2 & 77.7 & 33.3 & 36.5 & 25.2 & 25.7 \\   \midrule
      \multirow{8}{*}{OPT} & Null & "" & 50.4 & 71.7 & 25.4 & 29.4 & 18.3 & 18.8  \\ 
          & P-Tuning & - & 56.4 & 75.5 & 29.4 & 33.0 & 21.6 & 22.1  \\
       & RL-Prompt & "\textit{questions answers key question defining}" & \underline{59.2} & \underline{76.7} & \underline{31.5} & \underline{34.8} & \underline{23.4} & \underline{23.9}  \\
       & Manual & "\textit{Please write a question based on this passage}" & 55.3 & 74.6 & 28.7 & 32.4 & 21.1 & 21.6  \\ 
       & \multirow{3}{*}{Co-Prompt} & "\textit{Please tell that\ucr\space is the first question asked on Google for}" & \textbf{59.3} & \textbf{77.2} & \textbf{31.9} & \textbf{35.2} & \textbf{23.9} & \textbf{24.4}  \\
       & & "\textit{Please tell that\ucr\space is the question of}" & 58.8 & \underline{76.7} & 31.2 & 34.6 & 23.2 & 23.7  \\
       &  & "\textit{Please tell that\ucr\space is the first question to arise on}" & 58.3 & 76.0 & 31.0 & 34.3 & 23.1 & 23.5  \\  \midrule \midrule
      \multicolumn{2}{l|}{DPR} & - & 37.5 & 55.4 & 19.6 & 22.9 & 14.6 & 15.0  \\ \midrule
      \multirow{8}{*}{T0} & Null & "" & 51.5 & 68.0 & 27.8 & 30.9 & 20.9 & 21.3  \\ 
       & P-Tuning & - & 52.5 & 68.2 & 28.5 & 31.5 & 21.6 & 22.0  \\
      & RL-Prompt & "\textit{ question meaning difference meaning reality}" & 57.1 & 71.2 & 32.1 & 34.7 & 24.8 & 25.2  \\
       & Manual & "\textit{Please write a question based on this passage}" & \underline{57.7} & \textbf{72.0} & 32.2 & 34.9 & 24.8 & 21.4  \\ 
       & \multirow{3}{*}{Co-Prompt} & "\textit{Please send me some info on why or in detail,}" & \textbf{58.0} & \underline{71.7} & \textbf{32.7} & \textbf{35.3} & \textbf{25.3} & \textbf{25.7}  \\
       & & "\textit{Please send me some info on why or in detail about}" & 57.6 & 71.6 & \underline{32.5} & \underline{35.1} & \underline{25.1} & \underline{25.5}  \\
       &  & "\textit{Please send me some info on why or in detail on}" & 57.3 & 71.6 & 32.3 & \underline{35.1} & \underline{25.1} & \underline{25.5}  \\   \midrule
      \multirow{8}{*}{OPT} & Null & "" & 50.1 & 68.1 & 26.4 & 29.7 & 19.5 & 20.0 \\ 
        & P-Tuning  & - & 54.6 & 69.9 & 29.1 & 32.0 & 21.8 & 22.2   \\
       & RL-Prompt & "\textit{questions answers key question defining}" & \underline{56.3} & \underline{71.1} & \underline{31.1} & 33.8 & \underline{23.7} & \underline{24.1}  \\
       & Manual & "\textit{Please write a question based on this passage}" & 54.3 & 70.1 & 29.1 & 32.1 & 21.9 & 22.3 \\ 
       & \multirow{3}{*}{Co-Prompt} & "\textit{Please tell that\ucr\space is the first question asked on Google for}" & \textbf{56.4} & \textbf{71.3} & \textbf{31.4} & \textbf{34.2} & \textbf{24.1} & \textbf{24.5}  \\
       & & "\textit{Please tell that\ucr\space is the question of}" & \underline{56.3} & \underline{71.1} & \underline{31.1} & \underline{33.9} & \underline{23.7} & \underline{24.1} \\
       &  & "\textit{Please tell that\ucr\space is the first question to arise on}" & 55.8 & 70.6 & 30.8 & 33.5 & 23.5 & 23.9 \\  
     \bottomrule
    \end{tabularx}
    \caption{
    Detailed results of LLM re-ranker with different prompts. The performance is evaluated with the three metrics at top-20 and top-100 documents. 
    }
    \label{tab:gen_prompts}
\end{table*}

We evaluate the performance of zero-shot re-ranker with various metrics at Top-20 and Top-100 documents, as shown in Table \ref{tab:gen_prompts}. Co-Prompt is ranked 1st or 2nd on every metric across all experiments. On the other hand, the manual prompt, optimized for NQ, records inferior performance over MS-MARCO. Also, other optimization methods, RL-Prompt and P-Tuning, fail to achieve the best record in all experiments. This shows that the optimal prompt for zero-shot re-ranker is made from our method, Co-Prompt.

In addition, when confirming qualitatively generated prompts, the outputs from Co-Prompt are similar to human language patterns compared to RL-Prompt. The keyword "\textit{question}" is included in most of the prompts generated by Co-Prompt. Considering that other optimization methods produce dense prompt embedding or ungrammatical gibberish, Co-Prompt suggests a new direction in which a prompt can function as a natural user interface to understand a black-box model.

\end{document}